# Methods, Forms and Safety of Learning in Corporate Social Networks

Svitlana Lytvynova, Oleksandr Burov

Institute of Information Technologies and Learning Tools, Kyiv, Ukraine
s_litvinova@i.ua, ayb@iitlt.gov.ua

**Abstract.** The paper discusses methods, forms and safety issues of social network usage for school students. Taking into consideration the growing interest of students to electronic communication in social networks (ESN), it was described their place in the information educational environment. There were made the classification of objects and use of ESN to help teachers and school authority to teach students in the corporate social network. The basic components of corporate social networks (CESN) were revealed: forms of learning activity (individual, group, and collective), forms of learning (quiz, debates, discussions, photo-story, essay contest, a virtual tour, mini design web quest, and conference video-lesson), and database. They were defined particular aspects of the use of certain forms for students training in ESN according to the type of social objects (messages, individual messages, video files, photos, audio files, documents, comments, and blitz-survey). Student safety when using ESN is discussed as well.

**Key Terms:** TeachingMethodology, TeachingProcess, KnowledgeManagementMethodology, QualityAssuranceMethodology, StandardizationProcess.

## 1 Introduction

Human potential development is accompanied by key and continuously increasing role of education that has challenge related to the creation of appropriate informational learning environment. To date, informatization of education is characterized by the use of innovative information and telecommunication technologies, cloud computing, mass media technology and virtual reality systems and philosophical understanding of the process of informatization in education and its social consequences [6].

The use of cloud computing, including cloud services, cloud-based learning environment (Honshu) Office 365 in the general secondary education gave impetus to learning mobility to ensure all participants in the educational process [5].

Today social networks become increasingly popular among teachers of secondary schools and students as a tool for communication. However, the use of social networks for teachers is not completely explored, and requires special analysis and synthesis, especially because of new features (opportunities, cognitive potential and possible hazards) of open networks usage for learning and teaching [1].

The main difference is that networks have a special potential for students' socialization. If traditional teaching is aimed at the transfer of certain knowledge and the development of the information-transforming intellect, then digital space (first of all, social networks) can develop social and cultural intellect in accordance with modern trends in children needs [2; 3]. Each kind of intellect contributes to the integral "successful intelligence" (after R. Sternberg), but corresponds the specific kind of relations of a human with the external world (Fig. 1).

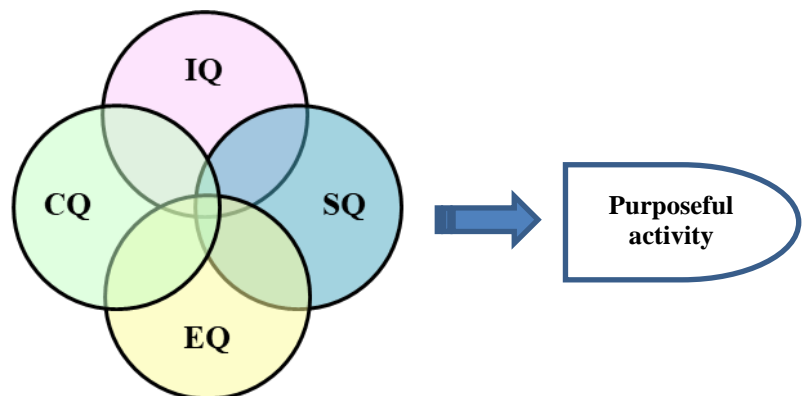

Fig.1 Successful intelligence as a form of integrated interaction of its structural components

Information and transforming intelligence (IQ) is associated with analysis' control and transformation of information, mutual adaptation of the individual and the environment ("Me - the World").

Social intelligence (SQ): personal development to the mutual adaptation of own characteristics and needs to the conditions of society ("Me – Others").

Emotional intelligence (EQ): coordination of individual responses of his inner world on the outer actions ("Me – Myself").

Cultural Intelligence (CQ) - analysis, conversion and use of own cultural opportunities (potential) and recognition of cultural peculiarities of other people ("Me – Mankind").

As a result of intellectual activity, a human creates his/her own and general information space, that is a set of results of semantic humanity activity.

The aim of the article is to analyze briefly methods, forms and safety issues of social network (SN) usage for school students.

## 2      Method

The analysis and synthesis of basic forms and methods of teaching students in social networks. It has been carried out survey of 250 teachers from Vinnytsia, Odesa, Khmelnytsky, Lugansk, Donetsk and Kyiv regions and analyzed the results. Besides, comparison of some data with results from the University of Phoenix (USA), conducted in 2015 to ascertain the state of social networking for teachers teaching students of secondary schools, provided.

## 3      Research findings

According to the research of foreign colleagues, 83% of 15-18-year-old students cannot do without high-speed Internet, a 88% use social networks every day [6; 7]. This trend in the use of ICTs should be taken into account in organizing learning activities of students as leaders of schools and teachers.

### 3.1      Forms of students training in corporate electronic social networks (CESN)

Innovative changes in secondary education and formation of a new Ukrainian school affect the development of professional and creative abilities of teachers in matters of creation of information learning environment (ILE) and the modern design of a lesson, including use of social networks (Fig. 2) especially electronic ones (ESN).

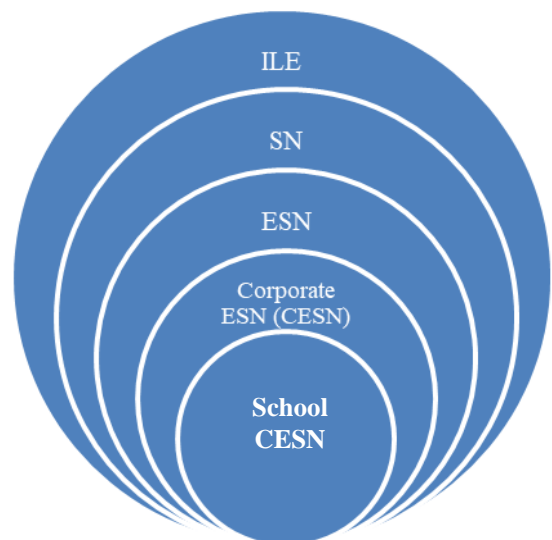

Fig. 2 Place of the corporate social network information and educational environment of a school

Studying the use of corporate social networks in general secondary education raises the question of classification of its main components: forms of education, social objects, shapes student learning and content databases.

The need in additional communication is a base of social networks [5]. There are two main areas of communication: indirect (communication not directly, but through a mediate link) and direct (personal). In social media, communication has the additional link: a social network [7]. Therefore, communication can be defined as indirect.

Social Network brings together actors, but it is needed specific factor of communication that is a social object (SO). Such a social object can be identified in all successful social networks, namely: video (YouTube), music (Last.fm), presentation (SlideShare), journalistic, reference articles (Wiki-Wiki), photos, projects (Scratch) etc. [1].

As social objects of ScoolCESN they could be considered: messages, personal messages, video, photos, audio, presentations, documents, messages, comments, blitz-survey.

In practice, such activities were recognized as successful: accommodation homework - documents; discussion of literary works - comments; narrative development - selected fragments; storytelling development - presentation; summarizing information on educational topics - surveys, etc.

The form of training needs the design of specific forms that provide conditions for effective training of students under the guidance of a teacher and realized the unity of content and technology education, which results in mastering subjects teaching knowledge, skills and development as a subject and key competencies.

The form of training is an external side of the educational process that reflects the way of students and teachers organization and is executed in certain order and mode, and depends on the number of students, the nature of the interaction of the learning process, level of autonomy, specific educational activities [1]. It involves organizing and setting up teacher interaction with students during their work with a certain content of learning materials [5]. Forms of education can be classified by the presence of students (online, offline) and the number of students: individual, group, collective [3; 4].

Communication between people is implemented in such structures: writing (mediated) and direct: individual; group; collective. In ESNs communication is mediated (subjects typing), but also in both a group, and mass. One can define the basic organizational forms of educational communication in social networks as individual, group and collective ones.

Individual form of learning activities of students in teacher-student social network assumes that every student gets a task for self-fulfillment, chosen for him according to his/her training and educational opportunities, and the teacher would give him/her an advice, tips and coordinate his/her activities. In SNs, students act in free pace, following only time limits pre-defined by the teacher. The student task performance carried out without communication with other students, foresees the development of individual cognitive and creative activities. The teacher can coordinate the training of each student according to his/her abilities.

Different educational groups can be created in CESN: virtual methodical association, project team, creative team, school council, partners and others. This grouping allows us to provide documents to share only in specific group of participants, to correspondence within the group and discuss topical issues, discuss documents, regulations and so on. A separate group can be formed from a school parents committee.

They need to be specified not only organizational learning forms, but forms of education of students as well. Forms of teaching students require thorough preparation teacher training materials and lesson plan. The most common forms are as follows: quiz, debates, photo story, essay contest, a virtual tour, mini-project, web-quest, conference.

Additional attention needs to be paid to classification of usage subjects of CESNs, such as: subject teachers, curators (class teachers), psychologists, social teachers, the administration of the institution (director, deputy director, secretary), school librarians, students, parents, representatives of-school education institutions, NGOs, donors, and district inspectors [6].

Another classification can be provided: by students age (elementary school, primary school, high school); by teachers age (20-35 years, 36-50 years, 51-60 years, 60+).

Classification of learning students used ESNs can be made on the following grounds as well:
− The location of the student. The school forms of education: lessons, work in workshops on near-the-school research station, laboratory and more. Out-of-form education, tour, home self-study, extracurricular activities at school; the company;
− The didactic purpose of students training: theoretical, practical, combined;
− Time of student learning. Time limit and after school, electives, subject groups, quizzes, competitions, subject evenings, etc.;
− The duration of stay in the student network, short message, detailed messages online communication.

### 3.2 Methods of education in ESNs

Teacher's expertise needs not only to know about his/her subject, but mustering teaching methods [5]. The network is already an effective learning tool as such, because teachers and students communication is realized through social objects (photos, videos, audio messages, presentations) that may already be an illustration or story, or statement of the problem for the lesson. Teachers only need to apply effective methods for collective, group or individual work with students [1].

We believe that the most successful methods of students teaching in secondary schools with CESN are those reflected a logic and perception of learning data. These methods of learning are inextricably linked with logical way of learning, namely, analysis, synthesis, comparison, generalization, specification, selection chief, classification, deduction, induction and control [4].

The choice of teaching methods depends on: the overall objectives of education and personal development; goals, objectives and content of the educational material for each lesson; features relevant content and methods of science, and that the subject of the topic being studied; features specific teaching methods of discipline; time spent on studying appropriate material; age characteristics of learners; level of preparedness (education and development); material equipment of the institution, including training equipment, visual aids, equipment; features and characteristics of the teacher, of his/her theoretical and practical expertise, personal qualities, pedagogical skills [1].

In Ukraine, over 2016 under the project "Cloud services in education", it was conducted a survey for teachers of secondary schools on the use of social media to enhance learning activities and interests of students in new forms of work, improving the educational process, communication skills and teamwork of students.

The survey involved 250 subject teachers from Vinnytsia, Odesa, Khmelnytsky, Lugansk, Donetsk and Kyiv regions [4; 5; 6].

It was revealed that the largest share of teachers who participated in the survey was 41-45 years. As a rule, their job experience reached 20 years, they have the highest category, and most of them have the title "Methodist".

Important data were found regards using social media for student learning. Compared with foreign counterparts', proportion of teachers who used social networks for learning was less and not more than 44%, but the proportion of those who do not use was less as well, 10.8%.

The ratio of teachers' attitude to SNs influences the development and formation of information-educational environment. When analyzing the survey results it was found that 66% of teachers are positive to social networks, 30.3% - neutral and 3.7% - negative, and this gives reason to believe that social networks can be effective means to support student learning. Teachers who have defined their attitude as "neutral" supports the idea of foreign counterparts, due to the lack of additional training, awareness of parents and lack of confidence in the safe use of social networks.

## 4   Discussion

Students' security issues in corporate networks remained because of many reasons: not enough awareness regards threats from the network and from ICT; lack of information regards psychology and psychophysiology of network users, especially in young age; lack of information regards ergonomics in ICT workplace etc.

During the survey, teachers were formulated question whether they secure social network for students. According to teachers, not all social networks are protected (Fig. 3).

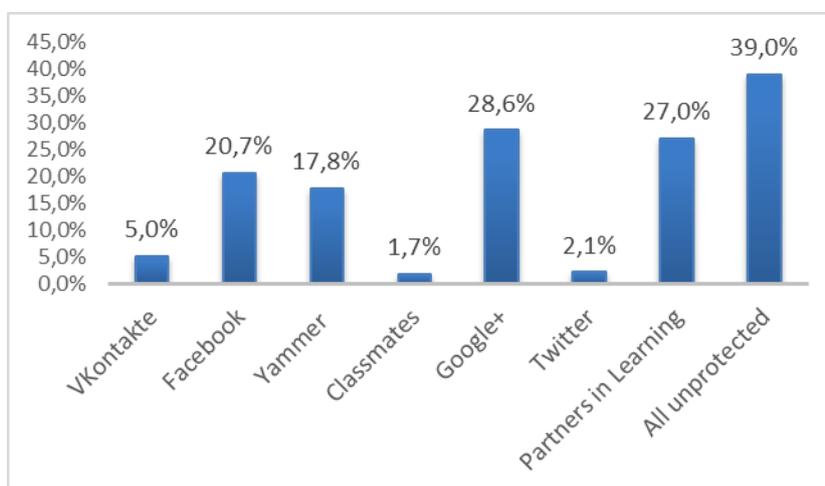

Рис. 4. Teachers' thought regards students' safety in ESNs

Nevertheless, 39% of teachers admit that social networks are not secure environment and it should work quite carefully.

Teacher answers the question concerning the effectiveness of training in social networks were as follows: Yes - 59.8% No - 8.7% could not decide - 31.5%. It was found that 31.5% of teachers indicated that they could not identify their attitude. That fact pointed the lack of criteria for evaluating the effectiveness of student learning in social networks. Unfortunately, it is necessary to highlight that not all teachers know and understand threats from the network enough clear [3] that is why they could not secure and have objective attitude to the network-centric learning appropriately.

There was a lively discussion with teachers how to improve the safety of students when using social networks. The panelists concluded that the use of corporate social networks (eg, Yammer) will enable students to be protected from contact with outsiders. However, responsibility for information located on the corporate network must carry both teachers and students. It is therefore advisable to conduct outreach on certain risks to intercourse in any social structure, especially in social networks.

Nevertheless, benefits of CESNs usage can be significant, if nurturing the necessary corporate safety culture: social networks allow accumulating social capital; social media can be a tool to implement community initiatives; they pro-vide users with marketing experience; social networks provide users' communi-cation temporal and special limitations; the economic effect of the use of social networks can be achieved by a reduction of portals' promotion; formation of social competence of students, including open reasoning, language culture and online communication.

## 5 Conclusions and outlooks

Corporate social networks play an important role to ensure communication and educational support for learning in the informational educational environment. For the organization of teaching students using social networking, it was proved essential for the implementation of the classification of objects and subjects of social networking, justified such social objects as messages, personal messages, video, photos, audio, presentations, documents, messages, comments, blitz poll.

As the most relevant, they were identified forms of learning (individual, group, collective) and singled forms of teaching students in social networks, namely, quiz, debates, discussions, photo-story, essay contest, a virtual tour, mini design web quest conference, video tutorials and others.

It is necessary to highlight the growing interest of students to communicate in social networks, teachers advisable to use their basic functionality to support the educational process in schools.

Identified methods of teaching students in social networks give impetus to the development of creative teamwork and building individual trajectory for students with special needs. The use of corporate social networks to discuss, evaluate, design work can serve as a basis for the development of creativity, openness, linguistic culture and online communication.

Special attention should be paid to student safety issues when using ESNs.